# On the Monetary Loss Due to Passive and Active Attacks on MIMO Smart Grid Communications

Ahmed El Shafie[†], Hamadi Chihaoui[⋆], Ridha Hamila[⋆], Naofal Al-Dhahir[†], Adel Gastli[⋆], Lazhar Ben-Brahim[⋆]
[†]University of Texas at Dallas, USA (e-mail: {ahmed.elshafie, aldhahir}@utdallas.edu).
[⋆]Qatar University (e-mail: chihaoui.hamadi@gmail.com, {hamila,adel.gastli,brahim}@qu.edu.qa).

*Abstract*—We consider multiple source nodes (consumers) communicating wirelessly their energy demands to the meter data-management system (MDMS) over the subarea gateway(s). We quantify the impacts of passive and active security attacks on the wireless communications system's reliability and security as well as the energy-demand estimation-error cost in dollars paid by the utility. We adopt a multiple-input multiple-output multi-antenna-eavesdropper (MIMOME) wiretap channel model. To secure the MIMO wireless communication system, the legitimate nodes generate artificial noise (AN) vectors to mitigate the effect of the passive eavesdropping attacks. In addition, we propose a redundant design where multiple gateways are assumed to coexist in each subarea to forward the consumers' energy-demand messages. We quantify the redundant designs impact on the communication reliability between the consumers and the MDMS and on the energy-demand estimation-error cost.

## I. INTRODUCTION

The smart grid (SG) revolutionizes the aging energy grid by enabling a more efficient and adaptive electric utility. In this context, two-way wireless communications is critical in connecting the different SG network nodes and ensuring efficient SG operation. Reliability and security are fundamental SG quality-of-service (QoS) requirements that impact both response time and information confidentiality for the communicating nodes.

Demand-side management (DSM) refers to a set of methods used by the utilities to manage the energy consumption efficiently using the existing SG infrastructure. The SG's wireless communications system must overcome the challenges of components failures and security attacks to maximize the utility's efficiency and reduce its response times.

### A. Related work

Reliability of the SG's communications system impacts how the energy generation and distribution will meet the consumers' energy requirements. The service area is generally divided into subareas each with multiple consumers. Each consumer is equipped with a smart meter to monitor the energy-consumption of electric devices. The reliability of the SG's wireless communications systems was investigated in several works, e.g., [1], [2] and the references therein. The authors of [1] analyzed the reliability of SG wireless communications systems that support DSM, where the utility implements an advanced metering infrastructure.

Recently, there has been increased research focus on SG security analysis, see, e.g., [3], [4] and the references therein. In [3], the authors provided an overview of SG security issues. SG security attacks can be classified into two categories: active attacks that attempt to disrupt the normal functionality of a network such as denial-of-service (DoS) attacks and passive attacks such as eavesdropping attacks [5] where the attackers attempt to decode and analyze the information exchanges to obtain important information about the legitimate transmitter-receiver nodes. Communication system secrecy from an information-theoretic perspective, also referred to as physical layer (PHY) security, was studied in the pioneering work of Wyner [6]. The system's PHY security is measured by the secrecy capacity of the link connecting the legitimate nodes, which represents the maximum rate of the legitimate transmitter-receiver pair with zero information leakage to the eavesdropping node.

Outage probability of the multiple-input multiple-output multi-antenna-eavesdropper (MIMOME) wiretap channels was analyzed in several works including [7], [8]. Reference [7] analytically optimized the secrecy rate of the MIMOME wiretap channel assuming statistical channel state information at the transmitter (CSIT). In [8], the system's secrecy was defined as a lower-bound on the minimum mean squared error between the transmitted and decoded data at the eavesdropping node. In [9], the authors investigated the impact of the wireless network's security and reliability on DSM operation in the SG for the single-input single-output (SISO) orthogonal frequency-division multiplexing (OFDM) wiretap channel. Assuming both active and passive attacks, the authors proposed a framework that connects the DSM and the outage probabilities of the wireless links. An artificial noise (AN) injection scheme was also proposed to enhance the security. In [10], the authors investigated the secrecy rate of MIMO wiretap channels and guaranteed a predefined signal-to-noise ratio (SNR) for successful decoding at the legitimate receiver.

### B. Contributions

Our main contributions are summarized as follows
- Unlike [9], we investigate the impact of the passive and active security attacks on the DSM operation for MIMO SG communications. We design the active attacks where the attacker sends a jamming signal to hurt the legitimate receivers without hurting the passive attacker (eavesdropper). In addition, we propose a new scheme to secure the legitimate system and derive the optimal transmit and receive filters that maximize the rates of the legitimate links.
- We analyze the achievable link rates under the proposed scheme and derive the necessary conditions on the number of antennas at each node to inject the AN signals.

- To further enhance the communications reliability, we propose a redundant design in which more than one gateway (Bob) is located in each subarea. In addition, we investigate the effect of this new design on both the PHY security and the utility's monetary loss due to energy-demand estimation errors.

*Notation:* Lower- and upper-case bold letters denote vectors and matrices, respectively. $\mathbf{I}_N$ stands for the identity matrix with size $N \times N$. $\mathbb{C}^{M \times N}$ is the set of complex matrices of size $M \times N$. $\mathbf{A}^*$ indicates the Hermitian transpose of matrix $\mathbf{A}$. $\mathbb{E}\{\cdot\}$ stands for statistical expectation. $\mathbf{0}_{M \times N}$ refers to the all zero matrix with size $M \times N$, $[\cdot]^+ = \max\{0, \cdot\}$ returns the maximum between the argument and zero, and $\binom{n}{K}$ denotes the $n$ choose $K$ operation.

## II. MAIN ASSUMPTIONS

Consider a wireless network composed of a set of $N$ consumers (Alices) communicating with their gateways (Bobs) which forward their energy-demand messages to a data aggregation unit (George) as shown in Fig. 1. We investigate the scenario where each subarea $q$ is assigned to $m \geq 1$ gateways $B_1^q, \cdots, B_m^q$. We assume dual-hop transmissions where each consumer transmits his data to Bob who decodes-then-forwards the data to George. We consider both passive and active security attacks on the legitimate communication system generated by an eavesdropper (Eve) and a jammer (Jimmy), respectively, over the two communication hops. The number of antennas at Alice, Bob, Eve, George and Jimmy are denoted by $N_A$, $N_B$, $N_E$, $N_G$ and $N_J$, respectively. The number of data streams transmitted from node $\nu$ to node $r$ is $\mathcal{N}_{\nu,r}$. Eve's instantaneous CSI is assumed to be unknown at the legitimate transmitters. As it will be explained shortly, passive attacks degrade the communication system's reliability since the legitimate transmitters will assign portions of their transmit powers to inject AN signals to secure their transmissions. Moreover, active attacks degrade reliability since Jimmy sends jamming signals to degrade the received signals at the legitimate receivers.

Since the network topology is well established and any change in the number of nodes in the network is rare, there is no congestion in the network [1], [9]. A time-division multiple-access (TDMA) scheme is assumed for all legitimate transmissions. Hence, each communication time slot is divided into $N$ non-overlapping time subslots (each with duration equal to the channel coherence time) and each consumer is assigned to one of these subslots. Each transmitting node encodes the data over $k$ communication subslots during each hour.

We assume a quasi-static flat-fading channel model with fixed channel coefficients during the coherence time. We denote the channel coefficient between the $k$-th antenna at node $\nu$ and the $l$-th antenna at node $r$ by $h_{\nu_k, r_l}$. The fading channel coefficients are assumed to be independent and identically distributed (i.i.d.) random variables. We denote the $n$-th Alice, the $i$-th Bob in subarea $q$, George, Eve, and Jimmy by $n$, $B_i^q$, $G$, $E$, and $J$, respectively. $\mathbf{H}_{\nu,r}$ refers to the channel matrix between node $\nu$ and node $r$. The thermal noise at the receiving node $r$ is modeled as a zero-mean additive white Gaussian noise (AWGN) random process with a variance of $\kappa_r$ Watts/Hz. We assume fixed-power transmissions with a transmit power spectral density of $P_\nu$ Watts/Hz for transmit node $\nu$.

### A. Energy-Demand Estimation-Error Cost

The DSM system operation is realized over two stages [1], [9]: the unit-commitment and the economic-dispatch stages. During the first stage, the utility reserves the energy supply based on the estimated energy demand of the consumers during a period of one hour. If the energy supply was under estimated, the utility will buy the energy difference in the economic-dispatch stage to prevent the under-supply situation. The energy demand-estimation cost of the $n$-th consumer whose energy data was not correctly decoded at MDMS is given by [9]

$$C(n) = p_{uc} \int_0^{\mu_n} (\mu_n - a) f_A^{(a)}(n)\, da + p_{ed} \int_{\mu_n}^{E_{\max}} (a - \mu_n) f_A^{(a)}(n)\, da \quad (1)$$

where $f_A^{(a)}$ is the probability density function of the actual energy consumption, $\mu_n$ is the mean energy demand of consumer $n$, $E_{\max}$ is the maximum energy consumption, and $p_{ed}$ and $p_{uc}$ are the energy prices in the economic-dispatch and the unit-commitment stages, respectively. The average energy-demand estimation-error cost is

$$\mathbb{E}\{C(n)\} = \sum_{n=1}^{N} C(n)\, P_{\text{outage}}^n \quad (2)$$

where $P_{\text{outage}}^n$ is the outage probability while transmitting the $n$-th Alice packet.

### B. Active Security Attacks

To jam the legitimate communications system, Jimmy injects a jamming (AN) signal to degrade the legitimate receivers' SNRs. The jamming signal precoding matrix is designed in a way that hurts the legitimate receivers only without degrading the eavesdropper's SNR.

## III. PROPOSED SECURE SCHEME AND ACHIEVABLE RATES

To secure the legitimate transmissions, each legitimate transmitter injects an AN signal along with its data signal. Each legitimate transmitting node (i.e., Alice or Bob) splits its transmit power between data and AN transmissions. A fraction $0 \leq \theta \leq 1$ of the total power $P_\nu$ of transmitting node $\nu$ is assigned to data transmission while the remaining fraction of $(1-\theta)P_\nu$ is assigned to AN transmission.

### A. First Hop

Assuming that the information signal and AN signal precoding matrices at node $\nu \in \{n, B_i^q\}$ are given by $\mathbf{P}_\nu$ and $\mathbf{Q}_\nu$, respectively, the received signal vector at Bob is

$$\mathbf{y}_{B_i^q} = \mathbf{H}_{n,B_i^q}\mathbf{P}_n\mathbf{x}_n + \mathbf{H}_{n,B_i^q}\mathbf{Q}_n\mathbf{z}_n + \mathbf{H}_{J,B_i^q}\mathbf{Q}_J\mathbf{z}_J + \mathbf{n}_{B_i^q} \quad (3)$$

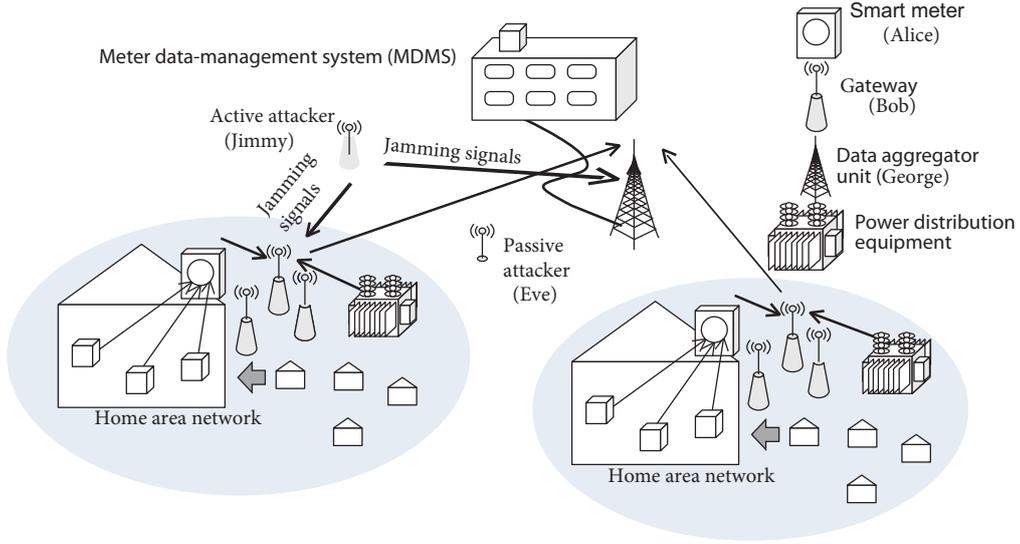

Fig. 1. System Model and security attacks.

where $\mathbf{x}_n \in \mathbb{C}^{\mathcal{N}_{A,B} \times 1}$ is the transmitted data vector by the $n$-th Alice, $\mathcal{N}_{A,B} \leq \min(N_A, N_B)$, $\mathbf{Q}_J$ is the used AN precoding matrix at Jimmy, $\mathbf{z}_J$ is Jimmy's AN vector, and $\mathbf{n}_{B_i^q} \in \mathbb{C}^{N_B \times 1}$ is the AWGN vector at Bob.

We assume a linear receive filter matrix $\mathbf{F}_{B_i^q}^*$ at Bob. The AN-precoding matrix at Alice should satisfy the following equation to cancel the AN at Bob

$$\mathbf{F}_{B_i^q}^* \mathbf{H}_{n,B_i^q} \mathbf{Q}_n = \mathbf{0}_{\mathcal{N}_{A,B} - (N_A - \mathcal{N}_{A,B})} \quad (4)$$

Let $\mathbf{W}_{B_i^q}$ be the covariance matrix of the interference vector $\mathbf{H}_{J,B_i^q} \mathbf{Q}_J \mathbf{z}_J + \mathbf{n}_{B_i^q}$. Then, $\mathbf{W}_{B_i^q} = \mathbf{H}_{J,B_i^q} \mathbf{Q}_J (\mathbf{H}_{J,B_i^q} \mathbf{Q}_J)^* + \kappa_{B_i^q} \mathbf{I}_{N_B}$. Bob uses the singular value decomposition (SVD) of the equivalent channel matrix to design a receive filter $\mathbf{\Psi}_{B_i^q}^*$ after applying a whitening filter $\mathbf{W}_{B_i^q}^{-\frac{1}{2}}$. Following the application of both filters, the received signal vector at Bob is given by

$$\mathbf{\Psi}_{B_i^q}^* \mathbf{W}_{B_i^q}^{-\frac{1}{2}} \mathbf{y}_{B_i^q} = \mathbf{\Psi}_{B_i^q}^* \mathbf{W}^{-\frac{1}{2}} \mathbf{H}_{n,B_i^q} \mathbf{P}_n \mathbf{x}_n + \mathbf{\Psi}_{B_i^q}^* \widetilde{\mathbf{n}}_{B_i^q} \quad (5)$$

where $\widetilde{\mathbf{n}}_{B_i^q} = \mathbf{W}_{B_i^q}^{-\frac{1}{2}}(\mathbf{H}_{n,B_i^q}\mathbf{Q}_n\mathbf{z}_n + \mathbf{H}_{J,B_i^q}\mathbf{Q}_J\mathbf{z}_J + \mathbf{n}_{B_i^q})$ and $\mathbf{F}_{B_i^q}^* = \mathbf{\Psi}_{B_i^q}^* \mathbf{W}_{B_i^q}^{-\frac{1}{2}}$ denotes the overall receive filter at Bob. From (4), $\mathbf{F}_{B_i^q}^* \mathbf{H}_{n,B_i^q} \mathbf{Q}_n = \mathbf{0}$, hence, $\mathbf{\Psi}_{B_i^q}^* \widetilde{\mathbf{n}}_{B_i^q} = \mathbf{F}_{B_i^q}^* \mathbf{H}_{J,B_i^q}\mathbf{Q}_J\mathbf{z}_J + \mathbf{n}_{B_i^q} + \mathbf{F}_{B_i^q}^* \mathbf{H}_{n,B_i^q} \mathbf{Q}_n = \mathbf{F}_{B_i^q}^* \mathbf{H}_{J,B_i^q}\mathbf{Q}_J\mathbf{z}_J + \mathbf{n}_{B_i^q}$.

Let $\mathbf{U}_{n,B_i^q} \mathbf{\Sigma}_{n,B_i^q} \mathbf{V}_{n,B_i^q}^*$ be the SVD of the equivalent channel matrix $\mathbf{W}_{B_i^q}^{-\frac{1}{2}} \mathbf{H}_{n,B_i^q}$. The data precoder $\mathbf{P}_n$ and the receive filter $\mathbf{F}_{B_i^q}$ are the $\mathcal{N}_{A,B}$ columns of $\mathbf{V}_{n,B_i^q}$ and $\mathbf{U}_{n,B_i^q}$, respectively, with the corresponding largest $\mathcal{N}_{A,B}$ non-zero singular values. The achievable rate at Bob is then given by

$$R_{n,B_i^q} = \log_2 \det \left( \mathbf{I}_{N_B} + p_n \widetilde{\mathbf{H}}_{n,B_i^q} \mathbf{P}_n (\widetilde{\mathbf{H}}_{n,B_i^q} \mathbf{P}_n)^* \mathbf{C}_{J,B_i^q}^{-1} \right) \quad (6)$$

where $p_n = \frac{\theta P_A}{\mathcal{N}_{A,B}}$, $\widetilde{\mathbf{H}}_{n,B_i^q} = \mathbf{W}_{B_i^q}^{-\frac{1}{2}} \mathbf{H}_{n,B_i^q}$ and $\mathbf{C}_{J,B_i^q} = \mathbf{W}_{B_i^q}^{-\frac{1}{2}} (\frac{P_J}{N_J - N_E} \mathbf{H}_{J,B_i^q} \mathbf{Q}_J (\mathbf{H}_{J,B_i^q} \mathbf{Q}_J)^* + \kappa_{B_i^q} \mathbf{I}_{N_B}) \mathbf{W}_{B_i^q}^{-\frac{1}{2}*}$.

The received signal vector at Eve is given by

$$\mathbf{y}_E = \mathbf{H}_{n,E} \mathbf{P}_n \mathbf{x}_n + \mathbf{H}_{n,E} \mathbf{Q}_n \mathbf{z}_n + \mathbf{H}_{J,E} \mathbf{Q}_J \mathbf{z}_J + \mathbf{n}_E \quad (7)$$

where $\mathbf{n}_E \in \mathbb{C}^{N_E \times 1}$ is the AWGN vector at Eve.

To cancel the AN at Eve, the AN-precoding matrix at Jimmy should satisfy

$$\mathbf{H}_{J,E} \mathbf{Q}_J = \mathbf{0}_{N_E \times (N_J - N_E)} \quad (8)$$

Note that the null space dimension is $\max\{N_J - N_E, 0\}$.

Finally, the achievable rate at Eve is given by

$$R_{n,E} = \log_2 \det \left( \mathbf{I}_{N_E} + p_n \mathbf{H}_{n,E} \mathbf{P}_A (\mathbf{H}_{n,E} \mathbf{P}_A)^* \mathbf{C}_{n-E}^{-1} \right) \quad (9)$$

where $\mathbf{C}_{n-E} = \frac{(1-\theta)P_A}{N_A - \mathcal{N}_{A,B}} \mathbf{H}_{n,E} \mathbf{Q}_n (\mathbf{H}_{n,E} \mathbf{Q}_n)^* + \kappa_E \mathbf{I}_{N_E}$.

*B. Second Hop*

The received signal vector at George is given by

$$\mathbf{y}_G = \mathbf{H}_{B_i^q,G} \mathbf{P}_{B_i^q} \mathbf{x}_{B_i^q} + \mathbf{H}_{B_i^q,G} \mathbf{Q}_{B_i^q} \mathbf{z}_{B_i^q} + \mathbf{H}_{J,G} \mathbf{Q}_J \widetilde{\mathbf{z}}_J + \mathbf{n}_G \quad (10)$$

where $\mathbf{x}_{B_i^q} \in \mathbb{C}^{\mathcal{N}_{B,G} \times 1}$ is Bob's transmitted data vector with $\mathcal{N}_{B,G} \leq \min(N_B, N_G)$, $\mathbf{Q}_J$ is the used AN precoding matrix at Jimmy, $\widetilde{\mathbf{z}}_J$ is Jimmy's AN vector during the second hop, and $\mathbf{n}_G \in \mathbb{C}^{N_G \times 1}$ is the AWGN vector at Eve.

Assume that a linear receive filter matrix $\mathbf{F}_G^* \in \mathbb{C}^{\mathcal{N}_{B,G} \times N_G}$ is used at George. Hence, to cancel Bob's AN vector at George's receiver, the $i$-th Bob in subarea $q$ sets

$$\mathbf{F}_G^* \mathbf{H}_{B_i^q,G} \mathbf{Q}_{B_i^q} = \mathbf{0}_{\mathcal{N}_{B,G} \times (N_B - \mathcal{N}_{B,G})} \quad (11)$$

Let $\mathbf{W}_G$ be the covariance matrix of the interference vector $\mathbf{H}_{J,G} \mathbf{Q}_J \widetilde{\mathbf{z}}_J + \mathbf{n}_G$. Then, $\mathbf{W}_G = \mathbf{H}_{J,G} \mathbf{Q}_J (\mathbf{H}_{J,G} \mathbf{Q}_J)^* + \kappa_G \mathbf{I}_{N_G}$. Bob uses the SVD of the equivalent channel matrix to design a linear receive filter $\mathbf{\Psi}_G^*$ after applying a linear whitening filter $\mathbf{W}_G^{-\frac{1}{2}}$. Then, the filtered received signal vector at Bob becomes equal to

$$\mathbf{\Psi}_G^* \mathbf{W}_G^{-\frac{1}{2}} \mathbf{y}_{B_i^q} = \mathbf{\Psi}_G^* \mathbf{W}_G^{-\frac{1}{2}} \mathbf{H}_{n,B_i^q} \mathbf{P}_n \mathbf{x}_n + \mathbf{\Psi}_G^* \widetilde{\mathbf{n}}_G \quad (12)$$

where $\widetilde{\mathbf{n}}_G = \mathbf{W}_G^{-\frac{1}{2}}(\mathbf{H}_{B_i^q,G}\mathbf{Q}_{B_i^q}\mathbf{z}_{B_i^q} + \mathbf{H}_{J,G}\mathbf{Q}_J\mathbf{z}_J + \mathbf{n}_G)$ and $\mathbf{F}_G^* = \mathbf{\Psi}_G^*\mathbf{W}_G^{-\frac{1}{2}}$ denotes the overall receive filter at George. From (11), $\mathbf{F}_G^*\mathbf{H}_{B_i^q,G}\mathbf{Q}_{B_i^q} = \mathbf{0}$, then $\mathbf{\Psi}_G^*\widetilde{\mathbf{n}}_G = \mathbf{F}_G^*(\mathbf{H}_{J,G}\mathbf{Q}_J\mathbf{z}_J + \mathbf{B}_G) + \mathbf{F}_G^*\mathbf{H}_{B_i^q,G}\mathbf{Q}_{B_i^q} = \mathbf{F}_G^*(\mathbf{H}_{J,B_i^q}\mathbf{Q}_J\mathbf{z}_J + \mathbf{n}_G)$. The data precoder $\mathbf{P}_{B_i^q}$ and the receive filter $\mathbf{F}_G$ are also derived based on the SVD of the equivalent channel matrix $\mathbf{W}_G^{-\frac{1}{2}}\mathbf{H}_{B_i^q,G}$.

The achievable rate at George is

$$R_{B_i^q,G} = \log_2 \det\left(\mathbf{I}_{N_G} + p_{B_i^q}\widetilde{\mathbf{H}}_{B_i^q,G}\mathbf{P}_{B_i^q}(\widetilde{\mathbf{H}}_{B_i^q,G}\mathbf{P}_{B_i^q})^*\mathbf{C}_{J-G}^{-1}\right) \quad (13)$$

where $p_{B_i^q} = \frac{\theta P_{B^q}}{\mathcal{N}_{B,G}}$ and $\widetilde{\mathbf{H}}_{B_i^q,G} = \mathbf{W}_G^{-\frac{1}{2}}\mathbf{H}_{B_i^q,G}$ and $\mathbf{C}_{J-G} = \mathbf{W}_G^{-\frac{1}{2}}\left(\frac{P_J}{N_J-N_E}\mathbf{H}_{J,G}\mathbf{Q}_J(\mathbf{H}_{J,G}\mathbf{Q}_J)^* + \kappa_G\mathbf{I}_{N_G}\right)\mathbf{W}_G^{-\frac{1}{2}*}$.

The received signal vector at Eve is given by

$$\mathbf{y}_E = \mathbf{H}_{B_i^q,E}\mathbf{P}_{B_i^q}\mathbf{x}_{B_i^q} + \mathbf{H}_{B_i^q,E}\mathbf{Q}_{B_i^q}\mathbf{z}_{B_i^q} + \mathbf{H}_{J,E}\mathbf{Q}_J\mathbf{z}_J + \mathbf{n}_E \quad (14)$$

where $\mathbf{n}_E \in \mathbb{C}^{N_E \times 1}$ is the AWGN vector at Eve. Hence, the achievable rate at Eve is

$$R_{B_i^q,E} = \log_2 \det\left(\mathbf{I}_{N_E} + p_{B_i^q}\mathbf{H}_{B_i^q,E}\mathbf{P}_{B_i^q}(\mathbf{H}_{B_i^q,E}\mathbf{P}_{B_i^q})^*\mathbf{C}_{B,E}^{-1}\right) \quad (15)$$

where $\mathbf{C}_{B,E} = \frac{(1-\theta)P_{B^q}}{N_B - \mathcal{N}_{B,G}}\mathbf{H}_{B_i^q,E}\mathbf{Q}_{B_i^q}(\mathbf{H}_{J,B_i^q}\mathbf{Q}_{B_i^q})^* + \kappa_E\mathbf{I}_{N_E}$.

The instantaneous secrecy rate of the $n$-th Alice transmission to Bob during the $\ell$-th communication time slot is

$$R_{n-B_i^q,\text{sec}}^{\ell} = \left[R_{n-B_i^q}^{\ell} - R_{n-E}^{\ell}\right]^+ \quad (16)$$

The instantaneous secrecy rate of the $B_i^q$-George link during the $\ell$th communication time slot when the $i$-th Bob in subarea $q$ forwards the $n$-th Alice's data packet is given by

$$R_{B_i^q-G,\text{sec}}^{\ell} = \left[R_{B_i^q-G}^{\ell} - R_{B_i^q-E}^{\ell}\right]^+ \quad (17)$$

Hence, the secrecy rate of the $n$-th Alice's transmission is

$$R_{n,\text{sec}} = \min_{\ell \in 1,2,\ldots,k} \min\{R_{n-B_i^q,\text{sec}}^{\ell}, R_{B_i^q-G,\text{sec}}^{\ell}\} \quad (18)$$

Each Alice sends fixed-rate data packets in each hour. We denote by $\mathcal{R}$ the target secrecy rate. Given that the data vector is transmitted over $k$ communication time slots, the minimum of the $k$ rates should be greater than $\frac{\mathcal{R}}{k}$ for successful decoding. Otherwise, the system is said to be unsecured.

Assume that there exists $m \geq 1$ Bobs, denoted by $B_1^q, \cdots, B_m^q$ in each subarea $q$. The outage probability of the link between the $n$-th Alice and the $m$ Bobs is

$$P_{n-B^q} = 1 - \prod_{l=1}^{k}\left(1 - \prod_{i=1}^{m}\Pr\left\{R_{n-B_{q_i}}^{\ell} < \frac{\mathcal{R}}{k}\right\}\right) \quad (19)$$

where $1 - \prod_{i=1}^{m}\Pr\left\{R_{n-B_{q_i}}^{\ell} < \frac{R}{k}\right\}$ is the probability that the $l$-th message is at least successfully decoded at one Bob in the subarea $q$. The outage probability of the $q$-th Bob's transmission when he forwards the $n$-th Alice's data packet is

$$P_{B^q-G} = 1 - \prod_{l=1}^{k}\left(1 - \prod_{i=1}^{m}\Pr\left\{R_{B_{q_i}-G}^{\ell} < \frac{\mathcal{R}}{k}\right\}\right) \quad (20)$$

Thus, the outage probability of the $n$-th Alice transmission is

$$P_{\text{outage}}^{n} = 1 - (1 - P_{n-B^q})(1 - P_{B^q-G}) \quad (21)$$

Since Eve's instantaneous CSI is assumed to be unknown at the legitimate transmitting nodes, the gateway selection in a subarea is realized based on the legitimate links' CSI and rates. Hence,

$$\hat{i} = \operatorname*{argmax}_{i}\left\{\min\left\{R_{n,B_{q_i}}, R_{B_{q_i},G}\right\}\right\} \quad (22)$$

By using the encoding scheme over the available $k$ communications subslots, the secrecy outage probability of the $l$-th block (from the $k$ blocks of a data packet) is

$$P_{n,\text{sec}}^{l} = P_{n,\text{sec}}^{\text{block}} = \Pr\left\{\min\left\{R_{n-B_i^q,\text{sec}}^{l}, R_{B_i^q-G,\text{sec}}^{l}\right\} < \frac{\mathcal{R}}{k}\right\} \quad (23)$$

which is the probability that Eve can decode partially (or completely) the $n$-th Alice transmitted information from either the first hop or the second hop. Since the channels are i.i.d., the probability in (23) is independent of time and block indices. If $k - m$ blocks of a data packet experience a secrecy outage, which occurs with probability $\left(P_{n,\text{sec}}^{\text{block}}\right)^{k-m}$, Eve can decode a fraction $\frac{k-m}{k}$ of the $n$-th Alice's packet. Hence, the fraction of unsecured data transmitted from the $n$-th Alice to George is given by [9]

$$P_{n,\text{sec}} = \sum_{m=0}^{k}\frac{k-m}{k}\binom{k}{k-m}(P_{n,\text{sec}}^{\text{block}})^{k-m}(1 - P_{n,\text{sec}}^{\text{block}})^{m} \quad (24)$$

## IV. SIMULATIONS RESULTS

We consider an SG wireless network with $M = 200$ consumers. Each channel coefficient is assumed to be a complex circularly-symmetric Gaussian random variable with zero mean and unit variance. We assume the following simulation parameters: $\mathcal{N}_{A,B} = \min(N_A, N_B) = N_B = 3$ and $\mathcal{N}_{B,G} = \min(N_B, N_G) = N_G = 2$, $N_B = N_G + 1$, $N_A = N_B + 1$, $N_E = 3$ and $N_J = 4$. We also assume that the energy consumption of each consumer follows a normal distribution [1], [9] with mean of 3 kWh and standard deviation of 1.5 kWh. The maximum energy consumption of each consumer is set to 10 kWh.

Fig. 2 demonstrates the impact of the target secrecy rate $\mathcal{R}$ on the system's security. Increasing $\mathcal{R}$ degrades the security due to the increased secrecy outage events. Moreover, Fig. 2 reveals that the fraction of unsecured data decreases as the number of antennas at the legitimates nodes increases due to spatial diversity gains while it increases when increasing the active attacker antennas, denoted by $N_J$.

Fig. 3 quantifies the impacts of the target secrecy rate $\mathcal{R}$ and our proposed redundant design, where $m$ gateways (Bobs) are available in each subarea, on the outage probability. Increasing $m$ significantly outperforms the case of a single Bob (i.e., $m = 1$). For example, for $\mathcal{R} = 10$ bits/channel use, by increasing $m$ from 1 to 4, the outage probability is reduced by 75%.

The impacts of $\mathcal{R}$ and the number of gateways on the utility monetary loss in dollars due to energy-demand estimation errors are quantified in Fig. 4. In fact, increasing the number of gateways in each subarea results in a much lower cost since it mitigates channel fading by finding a gateway that maximizes the end-to-end rate and, hence, the energy-demands of the consumers will be delivered reliably to the MDMS. Moreover, increasing $\mathcal{R}$ increases the monetary loss until it saturates around \$9000 at high $\mathcal{R}$ as the outage probability approaches 1 at high $\mathcal{R}$.

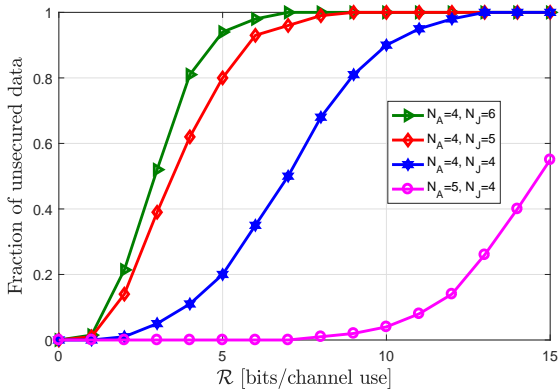

Fig. 2. Impact of $\mathcal{R}$ on system's security for different $N_A$ and $N_J$.

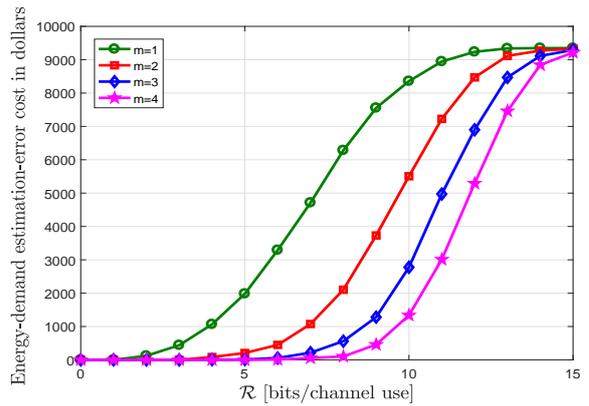

Fig. 4. Impacts of $\mathcal{R}$ and the number of gateways on the DSM error cost.

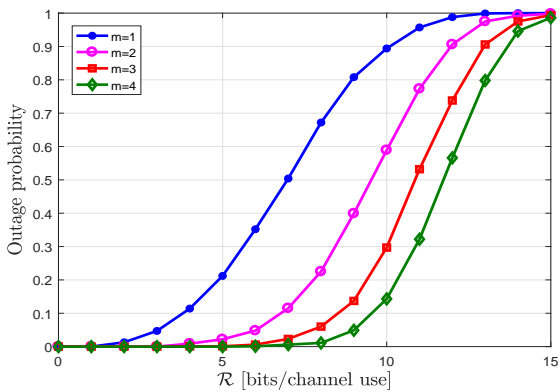

Fig. 3. Impacts of $\mathcal{R}$ and the number of gateways on the system's reliability.

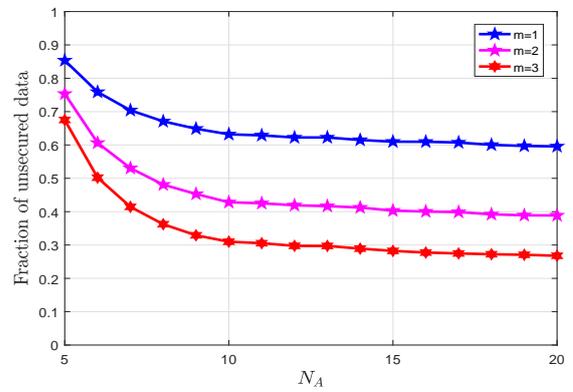

Fig. 5. Impacts of $N_A$ and the number of gateways on the system's security.

Finally, Fig. 5 demonstrates that as the number of gateways increases, the fraction of unsecured data decreases since the rates of the legitimate links will increase while the eavesdropping links will remain statistically the same. That is, in the gateway selection process, the relay that maximizes the legitimate link rate will be selected, which results in a non-decreasing rate for the legitimate system. This increases both security and reliability of the legitimate transmissions. Finally, increasing Alice's number of antennas $N_A$ enhances the system' security by decreasing the fraction of unsecured data since the legitimate link rate increases and more AN can be injected.

## V. CONCLUSIONS

We quantified the impact of the number of antennas at various nodes, the transmission data rate, and the gateway redundant design on the energy-demand estimation-error monetary loss for the utility and the system's reliability and security. We showed that by increasing the number of gateways from 1 to 4, the utility's monetary loss from \$8,357 to \$1,337 at target rate of 10 bits/channel use.